\documentclass[aps,prb,twocolumn,superscriptaddress,showpacs,preprintnumbers,amsmath,amssymb,floatfix]{revtex4}
\usepackage{graphicx}
\usepackage{dcolumn}
\usepackage{bm}
\usepackage{color}
\usepackage[latin1]{inputenc}
\graphicspath{{./figs/}}

\begin{document}

\title{Tunable effective masses of magneto-excitons in two-dimensional materials}

\author{A. Chaves} \email{andrey@fisica.ufc.br}
\affiliation{Universidade Federal do Cear\'a, Departamento de
F\'{\i}sica Caixa Postal 6030, 60455-760 Fortaleza, Cear\'a, Brazil}
\affiliation{Department of Physics, University of Antwerp, Groenenborgerlaan 171, B-2020 Antwerp, Belgium}
\author{F. M. Peeters}
\affiliation{Department of Physics, University of Antwerp, Groenenborgerlaan 171, B-2020 Antwerp, Belgium}

\begin{abstract}
Excitonic properties of Ge$_2$H$_2$ and Sn$_2$H$_2$, also known as Xanes, are investigated within the effective mass model. A perpendicularly applied magnetic field induces a negative shift on the exciton center-of-mass kinetic energy that is approximately quadratic with its momentum, thus pushing down the exciton dispersion curve and flattening it. This can be interpreted as an increase in the effective mass of the magneto-exciton, tunable by the field intensity. Our results show that in low effective mass two-dimensional semiconductors, such as Xanes, the applied magnetic field allows one to tune the magneto-exciton effective mass over a wide range of values.
\end{abstract}

\maketitle

\section{Introduction}

Chemical manipulation and addition of extra atoms in two-dimensional (2D) materials play an important role in expanding the already vast family of novel layered materials, allowing one to obtain compounds with different physical properties in a controllable manner.\cite{meng2015, mouri2013, fang2013, andreygapreview} For instance, with  hydrogenation, graphene can be engineered to exhibit semiconductor behavior, in contrast to its natural semi-metal character. \cite{whitenerjr, graphane} In this case, it becomes the so-called graphane\cite{graphane, graphane2, graphane3}, and its counterparts, known as silicene, germanene, and stanene, which have been coined the term Xenes,\cite{xenes, ni, fan, wang2016, brunetti2018, lyu} follow the same behavior, with their hydrogenized \cite{houssa2011, pham2015, bianco} versions being called Xanes. \cite{xanes, SG} 

Hydrogenation has been demonstrated to drastically modify optical properties of monolayer materials.\cite{SG, MF} In germanene, stanene or silicene, for example, the most significant opto-electronic properties mainly come from low-energy electrons around the K point, where the gap is zero, whereas germanane (Ge$_2$H$_2$), stanane (Sn$_2$H$_2$), and silicane (Si$_2$H$_2$) have their most relevant electronic states at the $\Gamma$-point, which exhibit gaps of 2.35 eV, 1.04 eV, and 3.83 eV, respectively, as obtained from GW calculations.\cite{SG, MF, Lu, cmr} Exciton binding energies obtained from Bethe-Salpether equations (BSE) are 290 meV and 390 meV for suspended Sn$_2$H$_2$ and Ge$_2$H$_2$, respectively.\cite{cmr} This demonstrates enhanced excitonic effects in their optical properties, which makes the investigation of excitons in these 2D hydrogenated materials very appealing.

In the presence of a perpericularly applied magnetic field $\vec{B} = B \hat{z}$, excitons undergo three different effects: (i) a Zeeman energy shift, which is linear with the field intensity $B$, (ii) a diamagnetic shift, that is positive and quadratic in $B$, and (iii) the so-called magneto-Stark shift, that is negative and quadratic in $B$. \cite{Glazov, Thomas, andreakou2016, karin2016} The latter is relevant when the momentum of the exciton center-of-mass $\vec{K}_{exc}$ is non-zero. Interestingly, this energy shift is quadratic in $K_{exc}$, thus representing a renormalization of the exciton effective mass. The effective mass enhancement of a magneto-exciton has been experimentally probed in coupled quantum wells of III-V semiconductor heterostructures. \cite{Lozovik2002} However, to the best of our knowledge, no magneto-exciton mass enhancement has been yet probed in atomically thin 2D semiconductors. This is due to the fact that this effect is hindered when effective masses and binding energies are high, which is the case with most 2D materials. However, Ge$_2$H$_2$ and Sn$_2$H$_2$ stand out as materials with very low effective masses, \cite{gosh2014, cmr} while the exciton binding energy in these materials can be controlled by the dielectric environment surrounding them. Thus, we may expect to observe strong control of the magneto-exciton effective mass in these materials by simply tuning the externally applied field. 

In this work, we theoretically investigate magnetic field effects on excitons in monolayer Ge$_2$H$_2$ and Sn$_2$H$_2$. Ge and Sn were chosen among all possible Xanes because they exhibit very low effective masses, in contrast to graphane and silicane,\cite{cmr, zolyomi2014} so the tunability of the magneto-exciton effective mass by means of the variation in the applied magnetic field can be significant. Our numerical results for excitons with zero center-of-mass momentum allows us to predict the diamagnetic shifts of excitons in these materials, whereas the calculated exciton energy dispersion provides information on the magneto-exciton mass. We demonstrate that the mass variation for small magnetic fields can be accurately predicted by using the in-plane electrical polarizability of the exciton. Therefore, as a parallel study, we also investigate the excitonic Stark effect for in-plane electric fields in Ge$_2$H$_2$ and Sn$_2$H$_2$.

\section{Theoretical Model}

In the presence of a magnetic field $\vec{B} = \vec{\nabla} \times \vec{A}$, the exciton Hamiltonian reads 
\begin{equation}
H_{exc} = \sum_{i = e,h}\left[\frac{\left(p^i_x + q A^i_x\right)^2}{2m_i}+\frac{\left(p^i_y + q A^i_y\right)^2}{2m_i}\right] + V_{eh}(|\vec{r}_e - \vec{r}_h|),
\end{equation}
where the vector potential $\vec{A^i}$ is written in the coordinates of the electron (hole) for $i = e(h)$, with $q = +e (-e)$. Due to the different screening environments in the substrate and capping layers, the electron-hole interaction potential has the Rytova-Keldysh form \cite{Rytova, Keldysh} 
\begin{equation}\label{eq.RytovaKeldysh}
V_{eh}(r) = -\frac{e^2}{2(\epsilon_1 + \epsilon_2)\rho_0}\left[H_0\left(\frac{r}{\rho_0}\right) - Y_0\left(\frac{r}{\rho_0}\right)\right],
\end{equation}
where $H_0$ and $Y_0$ are Struve and Neumann functions, respectively, $\rho_0 = \epsilon d/(\epsilon_1 + \epsilon_2)$ is the so-called screening length, with $d$ as the thickness of the semiconductor layer with dielectric constant $\epsilon$, and the dielectric constants for the surrounding top and bottom environments are $\epsilon_1$ and $\epsilon_2$, respectivelly.\cite{footnote} 

We now perform a transformation of coordinates, from the $(\vec{r}_e,\vec{r}_h)$ system to the relative $\vec{r}$ and center-of-mass $\vec{R}$ coordinates, along with a gauge transformation on the magnetic field. This leads to the Hamiltonian\cite{Lozovik2002, Muljarov, gorkov1968, lerner1980}
\begin{eqnarray}\label{eq.HamFinal}
H_{exc} = -\frac{\hbar^2}{2\mu}\nabla^2 + V_{eh}(r)  + \frac{\dot{\imath} \hbar eB}{2\eta}\left(x\frac{\partial}{\partial y} - y\frac{\partial}{\partial x}\right) \nonumber \\  + \frac{e^2 B^2 r^2}{8\mu}  +\frac{\hbar^2}{2M}K^2  + \frac{\hbar eB}{M}\left(xK_y - yK_x\right),
\end{eqnarray} 
where we assumed the vector potential in the Landau gauge $\vec A = \left(-By/2, Bx/2,0\right)$, and $\eta = (1/m_e - 1/m_h)^{-1}$, while $\mu = (1/m_e + 1/m_h)^{-1}$ and $M = m_e + m_h$ are the reduced and total effective masses, respectively.

\subsection{Stark shift and analytical approximation}

The Schr\"odinger equation for the Hamiltonian Eq. (\ref{eq.HamFinal}) cannot be solved analytically. Nevertheless, due to the low dielectric screening in such a thin layer material, the exciton binding energy in this system is higher than the energy scale of the corrections due to the last four terms of the Hamiltonian Eq. (\ref{eq.HamFinal}). Thus, these terms can be treated as perturbations upon a non-perturbed exciton Hamiltonian given by the first two terms in $H_{exc}$. Also, as we are dealing only with s-state excitons, the third term in Eq. (\ref{eq.HamFinal}), which couples the magnetic field to the exciton angular momentum, does not play a role, and can thus be removed. Consequences of the perturbative terms on the exciton energy levels can be predicted by simple arguments: notice that the terms that lead to magnetic field effects on the center-of-mass momentum in the Hamiltonian Eq. (\ref{eq.HamFinal}) are easily linked to the potential due to an effective in-plane electric field, $e\vec F = \hbar eB/M(-K_y, K_x, 0)$. The effect of an in-plane electric field on the exciton energy is well known \cite{Stark}: as in a hydrogen atom, the field leads to a quadratic Stark shift, i.e. the exciton energy will increase by $E_{Stark} \approx \alpha/2(F_x^2 + F_y^2)$, where $\alpha$ is the in-plane polarizability. Notice, however, that the effective electric field in this case is proportional to the center-of-mass momentum, therefore, the magnetic Stark shift reads 
\begin{equation}\label{eq.magnetostark}
E_{MStark} \approx -\frac{\hbar^2 B^2\alpha}{2M^2} K^2 ,
\end{equation}
This magnetic field induced analog of the Stark effect has been investigated before in the context of moving atoms, where it has been coined the term motional Stark effect, \cite{herold1981two, blumberg1979theory} as well as in bulk semiconductors, where it has been called magneto-Stark effect.\cite{Glazov, Thomas} The quadratic term in $\vec{K}$ in Eq. (\ref{eq.magnetostark}) couples to the kinetic energy of the center-of-mass in Eq. (\ref{eq.HamFinal}), which effectively becomes
\begin{equation}\label{eq.magnetostark2}
\frac{\hbar^2K^2}{2M} \rightarrow \frac{\hbar^2}{2M \left(1 - \frac{\alpha B^2}{M}\right)^{-1}}K^2 ,
\end{equation}
which allows one to conclude that the external magnetic field couples to the exciton by changing its total effective mass. Similar magnetic field dependent effective mass of magneto-excitons has been experimentally probed about two decades ago \cite{Lozovik2002} in the context of semiconductor quantum wells, but, to our knowledge, effective mass changes in magneto-excitons are yet to be experimentally observed in any recently discovered 2D materials. Our calculations suggest that materials with very small exciton mass, such as the monolayer Xanes investigated further on in this paper, are ideal for the observation of the magnetic field induced exciton mass renormalization.

\section{Results and Discussion}

In what follows, we provide theoretical predictions of the diamagnetic shifts and magnetic field dependent effective masses of magneto-excitons in the monolayer Xanes Ge$_2$H$_2$ and Sn$_2$H$_2$. Effective masses for electrons and holes in Ge$_2$H$_2$ (Sn$_2$H$_2$) are 0.050$m_0$ (0.020$m_0$) and 0.090 $m_0$ (0.030$m_0$), respectively.  \cite{cmr} The effective dielectric constant $\epsilon$ of each monolayer is obtained from DFT results: calculations for the ground state exciton energy based on the Bethe-Salpeter equation (BSE) for suspendend monolayer Ge$_2$H$_2$ (Sn$_2$H$_2$) yield $E_{1s}$ = 390 meV (290 meV). However, more simple calculations based e.g. on the effective mass approximation with the Rytova-Keldysh potential Eq. (\ref{eq.RytovaKeldysh}) have been demonstrated to provide fairly accurate binding energies, especially for the exciton ground state. \cite{berkelbach, henriques, andreygapreview, andreypolaritonreview, latini} Therefore, assuming the screening length for such a suspended monolayer as $\rho_0 = \epsilon d/2$ and fitting $\epsilon d$ as to obtain the same ground state exciton binding energy as in BSE calculations, one obtains $\epsilon d = 48.5 \epsilon_0$\AA\, ($31 \epsilon_0$\AA\,) for monolayer Ge$_2$H$_2$ (Sn$_2$H$_2$). 

Actual monolayer samples are usually laid over substrates (e.g. bulk BN or SiO$_2$), rather than being suspended. This is taken into account simply by properly changing $\epsilon_1$ and $\epsilon_2$ in the screening lengh in Eq. (\ref{eq.RytovaKeldysh}). We observe that the ground state exciton binding energies in monolayer Ge$_2$H$_2$ (Sn$_2$H$_2$) on a BN substrate or encapsulated by BN are 128.3 (68.4) meV and 62.8 (29.0) meV, respectively. These values are an order of magnitude lower than those observed for other 2D semiconductors, such as monolayer black phosphorus \cite{AndreyBP, louie, castellanos} and transition metal dichalcogenides (TMDs).\cite{berkelbach, andreypolaritonreview, latini} This is a consequence of the much smaller effective masses for electrons and holes in Xanes.

\subsection{Diamagnetic shift}

\begin{figure}[!t]
\centering
\includegraphics[width = 1.0\linewidth]{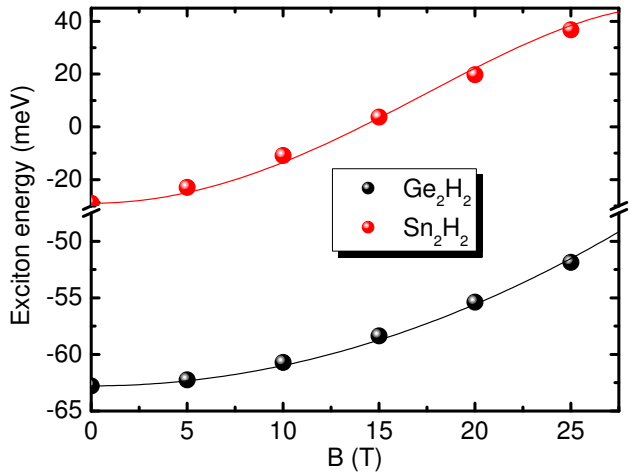}
\caption{Numerically calculated ground state exciton energies (symbols) in BN-encapsulated Ge$_2$H$_2$ and Sn$_2$H$_2$ as a function of the applied magnetic field. Solid lines are quadratic (black) and quartic (red) fittings to the numerical results.} \label{fig.diamagnetic}
\end{figure}

We numerically solved Schr\"odinger equation with the Hamiltonian Eq. (\ref{eq.HamFinal}), by diagonalizing the Hamiltonian discretized in a finite difference scheme. The exciton energies obtained by this procedure are shown as a function of the magnetic field in Fig. \ref{fig.diamagnetic}. As the magnetic field increases up to 25 T, the exciton energies with $K = 0$ increase by tens of meV. This is a much higher diamagnetic shift as compared to those observed in TMDs, \cite{donck, Stier2016, Stier2018, Zipfel2018, Goryca2019, Chen2019Luminescent} which is a consequence of the low binding energies in the BN-encapsulated Xanes considered here. 

In principle, the effect of the quadratic term in $B$ in Eq. (\ref{eq.HamFinal}) can be calculated in a perturbative way, since it is expected to be much smaller than the exciton binding energy. In this case, the diamagnetic shift in the exciton binding energy corresponds approximately to $\sigma B^2 = e^2 \langle r^2 \rangle B^2/8\mu$, where $\langle r^2 \rangle$ measures the broadening of the exciton wave function in real space, and the energy increases quadratically with $B$. This is indeed observed in Ge$_2$H$_2$, where the numerical results (symbols) in Fig. \ref{fig.diamagnetic} match well the quadratic fitting with $\sigma = 0.018$ meV/T$^2$ (solid black curve). However, for BN-encapsulated Sn$_2$H$_2$, where the exciton binding energy is much smaller and comparable to the diamagnetic shift, the usual quadratic fitting is no longer accurate enough and a quartic function $\sigma B^2 + \bar{\sigma} B^4$ is required for proper fitting of the numerical results. This is illustrated by the solid red curve in Fig. \ref{fig.diamagnetic}, given by a fitting function with a quadratic term $\sigma = 0.164$ meV/T$^2$ and a quartic term $\bar{\sigma} = -9\times 10^{-5}$ meV/T$^{4}$.

\subsection{Magneto-exciton dispersion}

\begin{figure}[!b]
\centering
\includegraphics[width = 1.0\linewidth]{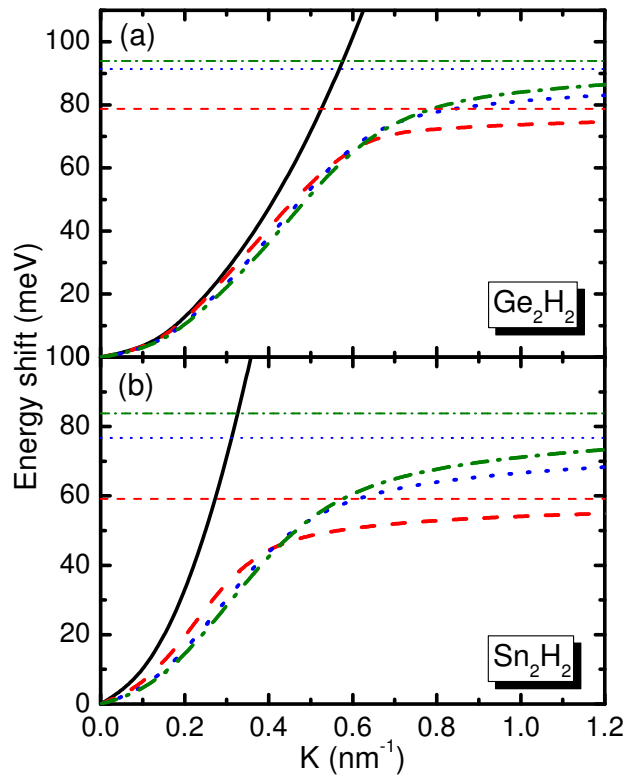}
\caption{Exciton energy shift as a function of the exciton center-of-mass momentum, for different values of the applied magnetic field $B$ = 0 T (black solid), 10 T (red dashed), 20 T(blue dotted), and 25 T (green dashed-dotted), for BN-encapsulated monolayer (a) Ge$_2$H$_2$ and (b) Sn$_2$H$_2$. Assymptotic values of each dispersion curve are represented by thin horizontal lines with the same color and line type.} \label{fig.dispersions}
\end{figure}

The exciton dispersions for BN-encapsulated monolayer Ge$_2$H$_2$ and Sn$_2$H$_2$ under several values of magnetic field are shown in Fig. \ref{fig.dispersions}. We verified that the effective mass model describes the exciton bands reasonably well in the range of center-of-mass momentum $K$ considered here. The exciton dispersion consists of two regimes: at low $K$, the curvature of the exciton dispersion strongly depend on the applied field. This constitutes the magneto-exciton regime, where the combination between the exciton and the applied field can be viewed as a quasi-particle whose effective mass is controlled by the field. At high center-of-mass momentum, on the other hand, the exciton band dispersion is flat, which represents a non-propagating quasi-particle. 

\begin{figure}[!b]
\centering
\includegraphics[width = 1.0\linewidth]{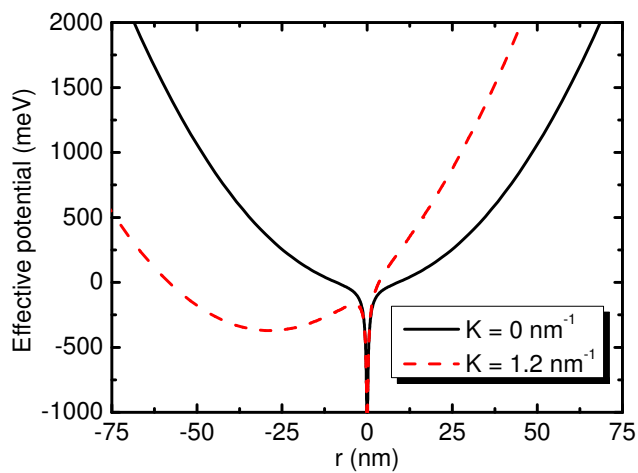}
\caption{Effective potential experienced by the exciton in BN-encapsulated monolayer Ge$_2$H$_2$ under an applied magnetic field $B$ = 25 T, assuming the center-of-mass momentum as $K$ = 0 nm$^{-1}$ or 1.2 nm$^{-1}$. Results for Sn$_2$H$_2$ are qualitatively similar, therefore, omitted here.} \label{fig.effectivepotential}
\end{figure}

The difference between the two regimes is better illustrated by the effective potential experienced by the exciton in the presence of the field, as shown Fig. \ref{fig.effectivepotential} for the case of BN-encapsulated monolayer Ge$_2$H$_2$ under a $B$ = 25 T applied magnetic field (results for Sn$_2$H$_2$ are qualitatively similar, therefore, omitted here). The effective potential results from the combination between the Rytova-Keldysh interaction potential $V_{eh}(r)$ and the  quadratic (fourth) and linear (sixth) terms in $r$ in Eq (\ref{eq.HamFinal}). The negative singularity at $r$ = 0 results from the interaction potential, whereas the parabolic profile for positive energies comes from the quadratic (diamagnetic) term. For vanishing center-of-mass momentum ($K$ = 0), these are the only terms involved and the effective potential is represented by the black solid curve in Fig. \ref{fig.effectivepotential}. In this case, the exciton is strongly confined in the negative cusp, while the parabolic potential slightly affects the exciton binding energy, as shown in Fig \ref{fig.diamagnetic}. However, as $K$ increases, a linear potential term in $r$ is added, thus shifting the position of the center of the parabolic potential, as illustrated by the red dashed curve in Fig. \ref{fig.effectivepotential} for $K = 1.2$ nm$^{-1}$. At high enough center-of-mass momentum, the ground state exciton wave function moves from the cusp to the bottom of the parabolic potential, where it ocupies the first Landau level of the system. In this case, the exciton band structure is flattened (see Fig. \ref{fig.dispersions}), the exciton velocity is zero and its mass is effectively infinite. 

In fact, in the limit of high magnetic fields and center-of-mass momenta, where the parabolic potential dominates and the electron-hole interaction can be taken as a perturbation, the exciton dispersion curves in Fig. \ref{fig.dispersions} asymptotically converge to $\hbar eB/2\mu - E_b$, where $E_b$ is the zero-field/zero-momentum exciton binding energy.\cite{Lozovik2002, gorkov1968, lerner1980} These asymptotic values, namely,  78.71 (59.13) meV, 91.38 (76.70) meV, and 93.89 (83.80) meV, respectively, for Ge$_2$H$_2$ (Sn$_2$H$_2$), are represented in Fig. \ref{fig.dispersions} by red dashed, blue dotted and green dashed-dotted thin horizontal lines, respectively.

\begin{figure}[!t]
\centering
\includegraphics[width = 1.0\linewidth]{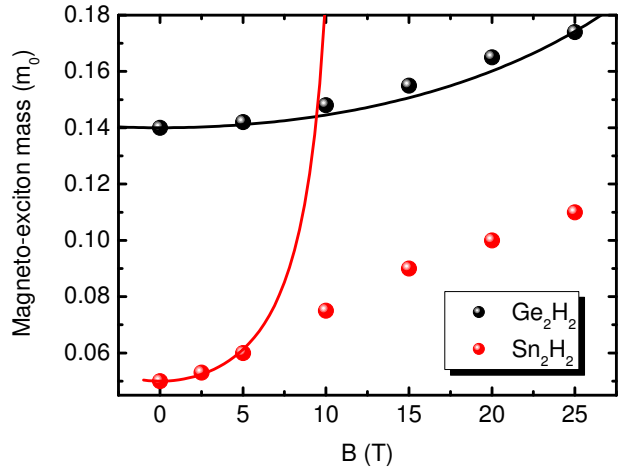}
\caption{Numerically calculated (symbols) magneto-exciton effective mass as a function of the applied magnetic field in BN-encapsulated Ge$_2$H$_2$ and Sn$_2$H$_2$. The curves are the approximate results using Eq. (\ref{eq.magnetostark2}). } \label{fig.masses}
\end{figure}

Due to the usually small momentum of the photon that excites the exciton, the exciton center-of-mass momentum is normally close to zero. Therefore, investigating the magneto-exciton effective mass in the vicinity of $K = 0$ is of major relevance. Numerical results obtained from the second derivative of the calculated magneto-exciton dispersions around $K = 0$ are shown as symbols in Fig. \ref{fig.masses} for BN-encapsulated monolayer Ge$_2$H$_2$ and Sn$_2$H$_2$. In the former, the magneto-exciton effective mass increases from 0.14 $m_0$ at zero field up to $\approx$0.175 $m_0$ at $B$ = 25 T, which represents a 25\% increase. For Sn$_2$H$_2$, due to the smaller effective mass and lower binding energy, the effect of the field is even more drastic: at $B$ = 25 T, the magneto-exciton effective mass is more than twice larger than its zero-field value. This demonstrates that the low effective mass Xanes investigated here, the exciton effective mass and, consequently, the exciton dynamics, can be efficiently controlled by means of an external magnetic field. 

As previously demonstrated in Eq. (\ref{eq.magnetostark2}), in-plane polarizabilities can be used to predict the magneto-exciton effective mass renormalization in low magnetic fields. Since, to the best of our knowledge, the Stark effect due to in-plane electric fields in monolayer Ge$_2$H$_2$ and Sn$_2$H$_2$ has not been investigated yet in previous papers, we have numerically calculated the Stark shifts for these materials. Polarizabilities found for monolayer Ge$_2$H$_2$ and Sn$_2$H$_2$ in the suspended case, as well as on a BN substrate or encapsulated by BN, are summarized in Table \ref{table}. They are, in general, an order of magnitude larger than those observed in black phosphorus and TMDs,\cite{AndreyBP,lucas2018,kamban,sun2020,haastrup2016, massicotte2018} which is also a consequence of the smaller effective masses and lower exciton binding energies.  

\begin{table}
\caption{Exciton polarizability for Ge$_2$H$_2$ and Sn$_2$H$_2$ in the suspended, on BN substrate, and BN-encapsulated cases, in units of 10$^{-4}$meV cm$^2$/kV$^2$.}
\label{table}\renewcommand{\arraystretch}{1.5}
\begin{ruledtabular}
\begin{tabular}{cccc}
Materials & Suspended & On substrate & Encapsulated \\\hline
Ge$_2$H$_2$ &-1.4518 &-7.2737 &-25.0198 \\
Sn$_2$H$_2$ &-4.4090 & -51.8487 &-208.4614  \\
\end{tabular}
\end{ruledtabular}
\end{table}

The magnetic field dependence of the magneto-exciton effective mass as predicted with Eq. (\ref{eq.magnetostark2}) and polarizabilities in Table \ref{table} is shown as solid lines in Fig. \ref{fig.masses}. Similar to the diamagnetic shift case in Fig. \ref{fig.diamagnetic}, the effective mass prediction is accurate only for Ge$_2$H$_2$ (black), where the exciton binding energy is stronger and the second order perturbation approach for the magnetic field contribution suffices. In contrast, for Sn$_2$H$_2$ (red), the prediction with Eq. (\ref{eq.magnetostark2}) fails for fields higher than 5 T, where higher order contributions in $B$ to the exciton energy, which are not accounted for in the derivation of Eq. (\ref{eq.magnetostark2}), play an important role.     

\section{Conclusions}

In summary, we have investigated the influence of a magnetic field on the effective mass of a magneto-exciton in monolayers of Ge$_2$H$_2$ and Sn$_2$H$_2$, also known as Xanes. In order to do so, we started our study by obtaining the exciton polarizabilities from the exciton Stark shift in the presence of an in-plane electric field. Polarizabilities are found to be an order of magnitude higher as compared to those observed in monolayer TMDs, due to the smaller effective masses of electrons and holes in Xanes. For the same reason, diamagnetic shifts are also shown to be stronger as compared to the ones in TMDs, being of the order of a few tens of meV for fields up to 25 T even for the exciton ground state. This, in principle, makes the diamagnetic shift of the 1s exciton state easier to be experimentally probed in Xanes. The magnetic field dependence of the effective mass of the magneto-exciton is numerically calculated, but an analytical prediction, involving the previously calculated exciton in-plane polarizability, is given as a good approximation for low fields, especially for Ge$_2$H$_2$. Our results show that the magneto-exciton effective mass in BN-encapsulated Sn$_2$H$_2$ can increase by more than a factor of two by the application of a field of 25 T. This allows for significant magnetic field control over the exciton dynamics in this material. 

\section*{Acknowledgements}

Discussions with G. O. de Sousa are gratefully acknowledged. This work was financially supported by the Brazilian Council for Research (CNPq), through the PRONEX/FUNCAP, Universal, and PQ programs, and by the Research Foundation-Flanders (FWO).


\begin{thebibliography}{apsrev}

\bibitem{meng2015}  Xiuqing Meng, Anupum Pant, Hui Cai,    Jun Kang, Hasan Sahin, Bin Chen, Kedi Wu, Sijie Yang, Aslihan Suslu, F. M. Peetersc, and Sefaattin Tongay, Nanoscale \textbf{7}, 17109 (2015) 

\bibitem{mouri2013} S. Mouri, Y. Miyauchi, and K. Matsuda, Nano Lett. \textbf{13}, 5944 (2013)

\bibitem{fang2013} Hui Fang, Mahmut Tosun, Gyungseon Seol, Ting Chia Chang, Kuniharu Takei, Jing Guo, and Ali Javey, Nano Lett. \textbf{13}, 1991 (2013)

\bibitem{andreygapreview} A. Chaves et al., npj 2D Materials and Applications \textbf{4}, 29 (2020)

\bibitem{whitenerjr} Keith E. Whitener Jr., Journal of Vacuum Science and Technology A \textbf{36}, 05G401 (2018)

\bibitem{graphane} Jorge O. Sofo, Ajay S. Chaudhari, and Greg D. Barber, Phy. Rev. B 75, 153401 (2007).

\bibitem{graphane2} M. Z. S. Flores, P. A. S. Autreto, S. B. Legoas, and D. S. Galvao, Nanotechnology \textbf{20}, 465704 (2009)

\bibitem{graphane3} Chao Zhou, Sihao Chen, Jianzhong Lou, Jihu Wang, Qiujie Yang, Chuanrong Liu, Dapeng Huang, and Tonghe Zhu, Nanoscale Research Letters \textbf{9}, 26 (2014)

\bibitem{xenes} A. Molle, J. Goldberger, M. Houssa, Y. Xu, S. C. Zhang, and D. Akinwande, Nat. Mater. 16, 163 (2017)

\bibitem{ni} Zeyuan Ni, Emi Minamitani, Yasunobu Ando, Satoshi Watanabe, Phys. Chem. Chem. Phys. \textbf{7}, 19039 (2015)


\bibitem{fan} Yingcai Fan, Xiaobiao Liu, Junru Wang,  Haoqiang Aib, and Mingwen Zhao, Phys. Chem. Chem. Phys.\textbf{20}, 11369 (2018)

\bibitem{wang2016} Can Wang, Qinglin Xia, Yaozhuang Nie, Mavlanjan Rahman, and Guanghua Guo, AIP Advances \textbf{6}, 035204 (2016)

\bibitem{brunetti2018} Matthew N. Brunetti, Oleg L. Berman, and Roman Ya. Kezerashvili, Phys. Rev. B \textbf{98}, 125406 (2018)

\bibitem{lyu} Ji-Kai Lyu, Shu-Feng Zhang, Chang-Wen Zhang, Pei-Ji Wang, Annalen der Physik \textbf{531}, 1900017 (2019)
 

\bibitem{houssa2011} M. Houssa, E. Scalise, K. Sankaran, G. Pourtois, V. V. Afanas'ev, and A. Stesmans, Appl. Phys. Lett. \textbf{98}, 223107 (2011)  

\bibitem{pham2015} Anh Pham, Carmen J. Gil, Sean C. Smith, and Sean Li, Phys. Rev. B \textbf{92}, 035427 (2015).

\bibitem{bianco} Elisabeth Bianco, Sheneve Butler, Shishi Jiang, Oscar D. Restrepo, Wolfgang Windl, and Joshua E. Goldberger, ACS Nano, 7, 4414 (2013)

\bibitem{xanes} V. Z\'olyomi, J. R. Wallbank, and V. I. Fal'ko, 2D Mater. \textbf{1}, 011005 (2014)

\bibitem{SG} O. Pulci, P. Gori, M. Marsili, V. Garbuio, R. Del Sole, F. Bechstedt, EPL \textbf{98}, 37004 (2012)

\bibitem{MF} Mojde Fadaie, Daryoosh Dideban, O\"guz G\"ulseren, Appl. Phys. A 126, 460 (2020)

\bibitem{Lu} Pengfei Lu, Liyuan Wu, Chuanghua Yang, Dan Liang, Ruge Quhe, Pengfei Guan, and Shumin Wang, Scientific Reports \textbf{7}, 3912 (2017)

\bibitem{cmr} Sten Haastrup, Mikkel Strange, Mohnish Pandey, Thorsten Deilmann, Per S. Schmidt, Nicki F. Hinsche, Morten N. Gjerding, Daniele Torelli, Peter M. Larsen, Anders C. Riis-Jensen, Jakob Gath, Karsten W. Jacobsen, Jens J. Mortensen, Thomas Olsen, Kristian S. Thygesen, 2D Materials \textbf{5}, 042002 (2018)

\bibitem{Glazov} A. Farenbruch, J. Mund, D. Fr\"ohlich, D. R. Yakovlev, M. Bayer, M. A. Semina, and M. M. Glazov
Phys. Rev. B \textbf{101}, 115201 (2020)

\bibitem{Thomas} D. G. Thomas and J. J. Hopfield, Phys. Rev. \textbf{124}, 657 (1961)

\bibitem{andreakou2016} P. Andreakou, A. V. Mikhailov, S. Cronenberger, D. Scalbert, A. Nalitov, A. V. Kavokin, M. Nawrocki, L. V. Butov, K. L. Campman, A. C. Gossard, and M. Vladimirova, Phys. Rev. B, \textbf{93}, 115410 (2016)

\bibitem{karin2016} Todd Karin, Xiayu Linpeng, M. M. Glazov, M. V. Durnev, E. L. Ivchenko, Sarah Harvey, Ashish K. Rai, Arne Ludwig, Andreas D. Wieck, and Kai-Mei C. Fu, Phys. Rev. B 94, 041201(R) (2016)

\bibitem{Lozovik2002} Yu. E. Lozovik, I. V. Ovchinnikov, S. Yu. Volkov, L. V. Butov, and D. S. Chemla, Phys. Rev. B \textbf{65}, 235304 (2002)

\bibitem{gosh2014} Ram Krishna Ghosh, Madhuchhanda Brahma, and Santanu Mahapatra, IEEE Transactions on Electron Devices \textbf{61}, 2309 (2014) 

\bibitem{zolyomi2014} V. Z\'olyomi, J. R. Wallbank, and V. I Fal'ko, 2D Materials \textbf{1}, 011005 (2014) 

\bibitem{Rytova} N. S. Rytova, Proc. MSU, Phys., Astron. 3, 30 (1967)

\bibitem{Keldysh} L. V. Keldysh, \textit{JETP Lett.} 29, 658 (1979)

\bibitem{footnote} Notice that the constants in this expression are slightly different as compared to those in F. Wu, F. Qu, and A. H. MacDonald, Phys. Rev. B \textbf{91}, 075310 (2015) and P. Cudazzo, I. V. Tokatly, and A. Rubio, Phys. Rev. B \textbf{84}, 085406 (2011). However, these definitions are all equivalent, differing only by their unit systems.

\bibitem{Muljarov} J. Wilkes and E. A. Muljarov, New J. Phys. \textbf{18}, 023032 (2016) 

\bibitem{gorkov1968} L. P. Gor'kov and I. E. Dzyaloshinskii, Sov. Phys. JETP \textbf{26}, 449 (1968) 

\bibitem{lerner1980} I. V. Lerner and Yu. E. Lozovik, Sov. Phys. JETP \textbf{51}, 588 (1980)

\bibitem{Stark} S. A. Empedocles and M. G. Bawendi, Science \textbf{278}, 2114 (1997)

\bibitem{herold1981two} H. Herold, H. Ruder, and G. Wunner, Journal of Physics B: Atomic and Molecular Physics \textbf{14}, 751 (1981)

\bibitem{blumberg1979theory} W. Blumberg, W. M. Itano, and D. J. Larson, Physical Review A \textbf{19}, 139 (1979)

\bibitem{berkelbach} Timothy C. Berkelbach, Mark S. Hybertsen, and David R. Reichman, Phys. Rev. B \textbf{88}, 045318 (2013)

\bibitem{henriques} J. C. G. Henriques and N. M. R. Peres
Phys. Rev. B \textbf{101}, 035406 (2020)

\bibitem{andreypolaritonreview} Tony Low, Andrey Chaves, Joshua D. Caldwell, Anshuman Kumar, Nicholas X. Fang, Phaedon Avouris, Tony F. Heinz, Francisco Guinea, Luis Martin-Moreno, and Frank Koppens, Nature Materials \textbf{16}, 182 (2017)

\bibitem{latini} Thomas Olsen, Simone Latini, Filip Rasmussen, and Kristian S. Thygesen, Phys. Rev. Lett. \textbf{116}, 056401 (2016)

\bibitem{AndreyBP} A. Chaves, T. Low, P. Avouris, D. \c{C}ak\i r, and F. M. Peeters, Phys. Rev. B 91, 155311 (2015)

\bibitem{louie} Diana Y. Qiu, Felipe H. da Jornada, and Steven G. Louie, Nano Lett. \textbf{17}, 4706 (2017)

\bibitem{castellanos} Andres Castellanos-Gomez, Leonardo Vicarelli, Elsa Prada, Joshua O. Island, K. L. Narasimha-Acharya, Sofya I. Blanter, Dirk J. Groenendijk, Michele Buscema, Gary A. Steele, J. V. Alvarez, Henny W Zandbergen, J. J. Palacios, and Herre S. J. van der Zant,
2D Materials \textbf{1}, 025001 (2014) 

\bibitem{donck} M. Van der Donck, M. Zarenia, and F. M. Peeters, Physical Review B \textbf{97}, 195408 (2018)

\bibitem{Zipfel2018} Jonas Zipfel, Johannes Holler, Anatolie A. Mitioglu, Mariana V. Ballottin, Philipp Nagler, Andreas V. Stier, Takashi Taniguchi, Kenji Watanabe, Scott A. Crooker, Peter C. M. Christianen, Tobias Korn, and Alexey Chernikov, Phys. Rev. B 98, 075438 (2018).

\bibitem{Stier2016} Andreas V. Stier, Kathleen M. McCreary, Berend T. Jonker, Junichiro Kono, and Scott A. Crooker, Nat. Commun. 7, 10643 (2016). 

\bibitem{Stier2018} A. V. Stier, N. P. Wilson, K. A. VElizhanin, J. Kono, X. Xu, and S. A. Crooker, Phys. Rev. Lett. 120, 057405 (2018).

\bibitem{Goryca2019} M. Goryca, J. Li, A. V. Stier, T. Taniguchi, K. Watanabe, E. Courtade, S. Shree, C. Robert, B. Urbaszek, X. Marie, and S. A. Crooker, \textit{Nature Communications} 10, 4172 (2019).

\bibitem{Chen2019Luminescent} Shao-Yu Chen, Zhengguang Lu, Thomas Goldstein, Jiayue Tong, Andrey Chaves, Jens Kunstmann, L. S. R. Cavalcante, Tomasz Wo\'zniak, Gotthard Seifert, D. R. Reichman, Takashi Taniguchi, Kenji Watanabe, Dmitry Smirnov, and Jun Yan, Nano Lett. 19, 2464 (2019).

\bibitem{lucas2018} L. S. R. Cavalcante, D. R. da Costa, G. A. Farias, D. R. Reichman, and A. Chaves, Phys. Rev. B \textbf{98}, 245309 (2018)

\bibitem{kamban} H. C. Kamban, Thomas G. Pedersen, and Nuno M. R. Peres, Phys. Rev. B \textbf{102}, 115305 (2020)

\bibitem{sun2020} Zheng Sun, Jonathan Beaumariage, Ke Xu, Jierui Liang, Shaocong Hou, Stephen R. Forrest, Susan K. Fullerton-Shirey, and David W. Snoke, Appl. Phys. Lett. \textbf{115}, 161103 (2019)

\bibitem{haastrup2016} Sten Haastrup, Simone Latini, Kirill Bolotin, and Kristian S. Thygesen, Phys. Rev. B \textbf{94}, 041401(R) (2016)

\bibitem{massicotte2018} Mathieu Massicotte, Fabien Vialla, Peter Schmidt, Mark B. Lundeberg, Simone Latini, Sten Haastrup, Mark Danovich, Diana Davydovskaya, Kenji Watanabe, Takashi Taniguchi, Vladimir I. Fal'ko, Kristian S. Thygesen, Thomas G. Pedersen, and Frank H. L. Koppens, Nature Communications \textbf{9}, 1633 (2018)


\end{thebibliography}
\end{document}